\documentclass[12pt]{iopart}
\usepackage{iopams}  
\usepackage{graphicx} 
\usepackage{cite}

\begin{document}
\newcommand*{\R}{\mathbb{R}}
\newcommand*{\N}{\mathbb{N}}
\newcommand*{\Q}{\mathbb{Q}}
\newcommand*{\Z}{\mathbb{Z}}
\newcommand*{\C}{\mathbb{C}}

\newcommand*{\be}{\begin{equation}}
\newcommand*{\ee}{\end{equation}}

\def\Varr{\mathrm{Var}\,}
\def\bE{{ \mathbb E}}
\def\ep{\varepsilon}
\def \ul {\underline}
\newcommand*{\Pv}{{\bf P}}
\newcommand*{\Ev}{{\bf E}}
\newcommand*{\Var}{{\bf Var}}
\newcommand*{\Cov}{{\bf Cov}}
\newcommand*{\Korr}{{\bf Korr}}

\newtheorem{conj}{Conjecture}
\newtheorem{theo}{Theorem}
\newtheorem{cor}{Corollary}
\newtheorem{ex}{Example}

\def\iA{\mathcal{A}}
\def\iH{\mathcal{H}}
\def\iM{\mathcal{M}}
\def\iE{\mathcal{E}}
\def\iB{\mathcal{B}}
\def\bF{\mathbf{F}}
\def\Tr{\mathrm{Tr}\,}
\def\<{\langle}
\def\>{\rangle}
\def\bbbc{{\mathbb C}}
\def\qed{\nobreak\hfill $\square$}

\title{Optimal parameter estimation of Pauli channels}

\author{L\'aszl\'o Ruppert$^{1,2}$, D\'aniel Virosztek$^2$ and Katalin Hangos$^1$}

\address{$^1$ Process Control Research Group, Computer and Automation Research Institute\\
H-1518 Budapest, POB 63, Hungary}

\address{$^2$ Department of Analysis, Budapest University of Technology and Economics,\\
H-1521 Budapest, POB 91, Hungary}

\ead{ruppertl@math.bme.hu, virosz89@gmail.com, hangos@scl.sztaki.hu}

\begin{abstract}

The optimal measurement configuration, i.e., the optimal input quantum state and measurement in the form of a POVM with two elements, is investigated in this paper for qubit and generalized Pauli channels. The {\it channel directions} are defined as the contracting directions of the channel with an arbitrary fixed basis of Pauli-matrices. In the qubit Pauli channel case with known channel directions it is shown that the optimal configuration that maximizes the Fisher information of the estimated parameters consists of three von Neumann measurements together with pure input states directed to the appropriate channel directions. Extensions of these results for generalized Pauli channels are also given. Furthermore, a computationally efficient estimation method is proposed for estimating the channel directions in the qubit Pauli channel case. The efficiency of this method is illustrated by simulations: it is shown that its variance is of order $1/N$, where $N$ is the number of used measurements.
\end{abstract}



\section{Introduction}

Accurate description of different quantum phenomena is a key issue in the 
 potential use of these phenomena in modern IT-technologies: communication, security, quantum computers, etc.  
The parameter estimation of quantum channels, that is commonly called quantum process tomography \cite{Dariano-2003}, 
plays a major role in quantum information processing. 
The direct way of quantum process tomography is performed by sending known quantum systems into the  
channel, and then estimating the output state. Experiment design is necessary if one wants to get efficient process tomography 
that consists of selecting the optimal input state, optimal measurement of the output state, and an efficient estimator of channel from the measurement data. 
In quantum mechanics the measurement has a probabilistic nature \cite{Nielsen-book,Petz-book}, therefore 
 many identical copies of the input quantum system are needed, and an estimator can be constructed by using some statistical considerations. 

The field of quantum process tomography is well established, an exhaustive description of possible tomography methods can be found in \cite{Mohseni-2008}. Within this field a relatively wide family of quantum channels described as Pauli-channels \cite{Nielsen-book} are of large theoretical and practical importance. The tomography of Pauli channels has a huge literature but - because of the hardness of the topic - the papers are mostly dealing with special cases. For example, in \cite{Sasaki-2002} the depolarization channel is examined. In the case of this one-parameter family of channels, the optimal measurement setup can be easily found. The optimal measurement is done via coupling an ancilla system, which is a common method in channel tomography \cite{Sarovar2006,Ji-2008,Mohseni-2008,Bendersky-2008}, because the tomographic efficiency can be improved via entanglement. 
At the same time, this is not necessary true in general \cite{Bschorr-2001}, so direct methods are widely spread too.

If one wants to find an optimal estimation method, the common solution is that some kind of Fisher information is maximized that  will result in minimal variance hence Cramer-Rao inequality. A similar problem appears in \cite{Dariano-2005}, where the discrimination of two different Pauli-channels is investigated. While the efficiency of the discrimination 
requires another quantity, but the optimal estimation method is the same as in standard tomography problems. 
Many recent experimental results have also proposed, that provide optimal estimation of the Pauli-channels in practice \cite{Young-2009,Branderhorst2009,Chiuri-2011}.

Each estimation method is constrained by the fact, that quantum channels must be completely positive and trace-preserving (CPTP), otherwise their description is physically not meaningful. Having estimated a non-physical map, one can construct from that an appropriate one \cite{Ziman-2005}. In the case of large number of measurements, however, the law of large numbers ensures that the estimated values will be close to the real channel parameters, which are obviously physically possible.

While most papers deal with qubit and one-parameter case, there are some exceptions. 
There are some papers investigating the estimation of multi-parameter channels\cite{Bendersky-2008,Young-2009,Nunn2010}, and the multidimensional case also appears in \cite{Fujiwara2003} and \cite{Nathanson-2007}. 
Although there are alternative ways for the generalization of Pauli-channels to multidimensional case, the above cited articles use a given basis and represent the Pauli-channels by a contraction in every direction. The paper \cite{Petz-2008} proposes  another way for the generalization of Pauli-channels:  the state space is divided on complementary subalgebras, and the contraction is applied on these subalgebras. This way the completely positiveness of mapping can be derived easily. 
This latter approach is followed here when dealing with generalized Pauli-channels.

\medskip
The aim of this work is to find the optimal measurement configuration consisting of an optimal 
quantum input state and a measurement POVM that provide a statistically optimal estimate 
of the channel parameters. For the sake of simplicity, in this work we will only use POVMs with two elements. 
An earlier engineering approach for this problem, that is called an experiment design problem in system identification, has been 
reported in \cite{Ballo-2011}, where the problem of optimal estimation of 
Pauli-channel parameters was investigated using convex optimization methods. 
In contrast to this, purely statistical methods are applied in this paper that enable us to obtain results analytically. 

The paper is organized as follows. 
In Section 2 we introduce the necessary notions to understand the rest of the article. 
In Section 3 and 4 we are assuming that the channel directions are known, and we give the optimal measurement configuration for both the qubit and the generalized Pauli channel case. 
In Section 5 the method for estimating the channel directions are described 
and analyzed. Finally, conclusions are drawn.


\section{Preliminaries}

\subsection{Quantum states and directions}

The quantum state of a finite $n$-dimensional system can be described using $\rho\in M_n(\bbbc)$ fulfilling the conditions:
\begin{equation}\label{E:state}
\rho \ge 0,  \quad \Tr(\rho)=1.
\end{equation}
In the special case of qubits (i.e., $n=2$), we will use Bloch parametrization:
\begin{equation}\label{E:2x2}
\rho_{\theta}= \frac{1}{2}\left[\begin{array}{cc}
1+\theta_3 & \theta_1 - i \theta_2 \\
\theta_1 + i \theta_2 & 1-\theta_3 \end{array}\right]=
\frac{1}{2}\left(I + \sum_{i=1}^3 \theta_i \sigma_i \right)\ ,
\end{equation}
where the so-called Pauli matrices
\begin{equation}\label{E:pauli}
\fl \quad I=\sigma_0=\left[\begin{array}{cc}
1 & 0\\
0 & 1  \end{array}\right]
,~
\sigma_1=\left[\begin{array}{cc}
0 & 1 \\
1 & 0 \end{array}\right]
,~
\sigma_2=\left[\begin{array}{cc}
0 & -i \\
i & 0 \end{array}\right]
,~
\sigma_3=\left[\begin{array}{cc}
1 & 0 \\
0 & -1\end{array}\right]
\end{equation}
are used as a basis in $M_2(\bbbc)$. From (\ref{E:state}) it  follows, that all states can be described with the Bloch vector $\theta=(\theta_1,\theta_2,\theta_3)\in \mathbb{R}^3$ fulfilling the condition
\begin{equation}\label{E:2cond}
\theta_1^2+\theta_2^2+\theta_3^2 \le 1.
\end{equation}

In the $n$-dimensional case let us use an arbitrary self-adjoint orthonormal basis, $v_i$ instead of Pauli-matrices 
that results in the parametrization
\begin{equation}\label{E:state_par}
\rho_{\theta}= \sum_{i=0}^{2^n-1} \theta_i v_i=\frac{I}{n}+\sum_{i=1}^{2^n-1} \theta_i v_i,
\end{equation}
where we used $v_0=\frac{1}{\sqrt{n}}I$. Then from (\ref{E:state}) it follows that $\theta_0=\frac{1}{\sqrt{n}}$, but the positivity condition can not be written in a simple, explicit form \cite{Petz-book}.

If $n=2^k$ then we can get the basis from the tensor product of the Pauli-matrices:
$\frac{1}{\sqrt{n}}\otimes_{i=1}^k \sigma_{j_i}$, where $\forall i:~ j_i \in \{0,1,2,3\}$.

Since density matrices can be described by Bloch vectors, we can consider the quantum states as elements of $\mathbb{R}^{2^n-1}$. The \textit{direction of a Bloch vector} is defined as the direction of this vector in $\mathbb{R}^{2^n-1}$ with the basis of the Pauli-matrices. 

\subsection{Quantum measurements}

Quantum measurements are represented mathematically by 
positive operator-valued measures (POVMs). 
These can be described as a set of $n \times n$ matrices satisfying: 
\begin{equation}\label{E:POVM}
\forall \alpha:~ M_\alpha \ge 0 \quad \textrm{and} \quad \sum_{\alpha} M_\alpha=I
\end{equation}
We will use only POVMs with two elements: $\{M,I-M\}$.
We can parametrize the POVM element $M$ in the same basis as we did in (\ref{E:state_par}): 
\begin{equation}\label{E:povm_par}
M= \sum_{i=0}^{n-1} m_i v_i=m_0\sqrt{n} \left(\frac{I}{n}+\sum_{i=1}^{n-1} \frac{m_i}{m_0\sqrt{n}} v_i\right)=m_0\sqrt{n} \left(\frac{I}{n}+\sum_{i=1}^{n-1} m_i^* v_i\right)
\end{equation}
thus $M$ can be represented with a Bloch vector $m=(m_0,m_1,\dots,m_{n-1}) \in \mathbb{R}^n$. The state space of such POVMs has similar structure as the space of quantum states (cf. $\theta_i$ and $m_i^*$), the difference is that $m_0$ is not fixed. The maximal value for $m_0$ is $\sqrt{n}$, in that case we have $\forall i>0:~m_i=0$ and so $M=I$. An arbitrary POVM fulfills the conditions in (\ref{E:POVM}) if all eigenvalues of $M$ are in the interval $[0,1]$. If all eigenvalues are 0 or 1, then $M$ is a projection and the POVM is called a von Neumann measurement.

The probability of measuring $M$ is
\begin{equation} \label{eq:prob}
p=\Tr(\rho_\theta M)=\sum_{i=0}^{n-1} m_i \theta_i =\frac{m_0}{\sqrt{n}}+ \sum_{i=1}^{n-1} m_i  \theta_i =\frac{m_0}{\sqrt{n}}+ d,
\end{equation}
where we used the orthonormality of basis ($\<v_i,v_j\>=\delta_{i,j}$) and the abbreviation
\begin{equation} \label{eq:d}
d= \sum_{i=1}^{n-1} m_i  \theta_i.
\end{equation}

\subsection{Quantum Pauli channels}

Quantum Pauli channels form a well-known and wide family of quantum channels in the 2-dimensional (qubit) case. Their usual definition implements contractions in the directions of the Pauli-matrices (basis vectors), but we define it with arbitrary directions.

Let $v_i$ be a self-adjoint orthogonal basis of $M_2(\bbbc)$, then the effect of the channel on an input state $\rho_{\theta}$ can be described as
\begin{equation}\label{eq:2x2out}
\rho_{\theta}=\frac{1}{2}\left(I + \sum_{i=1}^3 \theta_i v_i \right)
\longrightarrow \mathcal{E}(\rho_\theta)=\frac{1}{2}\left(I + \sum_{i=1}^3 \lambda_i \theta_i v_i \right).
\end{equation}
Hence the image of the Bloch ball will be an ellipsoid with semi-axes corresponding to $v_i$ directions and with length $\lambda_i$. 


A possible way of defining generalized Pauli channels is proposed in \cite{Petz-2008}. Assume that $M_n$ contains  complementary subalgebras 
$\iA_1,\iA_2, \dots \iA_{N_s}$, where all $\iA_i$-s have the same dimension $s$. They are complementary in the sense that their traceless parts are orthogonal:
$$
A_i -\frac{\Tr A_i}{n} I ~\perp~  A_j-\frac{\Tr A_j}{n} I, \quad \forall A_i \in \iA_i, A_j \in \iA_j, \quad i \ne j
$$
The above subalgebras represent now the channel directions. Then by simple dimension counting we get:
$$
(s-1) N_s=n^2-1
$$

In order to describe the effect of the generalized Pauli channel, 
the trace preserving projection $E_i:M_n \to \iA_i$ is defined first which is usually
called conditional expectation.
Then, for  any input $A\in\mathcal{B}(\mathcal{H})$ in the form 
$$
 A=- \frac{(N_s-1) \Tr A}{n} + \sum_{i=1}^{N_s} E_i (A)~,
$$
the output of the generalized Pauli channel is
\begin{equation}\label{E:gpc}
\iE(A)=\left(1-\sum_{i=1}^{N_s}\lambda_i\right)
\frac{\Tr A}{n}I +\sum_{i=1}^{N_s}\lambda_iE_i(A)
\end{equation}
The conditions for complete positivity and trace preserving property are 
\begin{equation} \label{eq:genPauli_inequ}
1+n \lambda_i \ge \sum_j \lambda_j \ge - \frac{1}{n-1}\ ~,~~~| \lambda_i| \leq 1 .
\end{equation}

If the subalgebras are generated by some elements of the basis $\{v_i\}$ in (\ref{E:state_par}), then the channel can be described as:
\begin{equation}\label{ch_gen}
\mathcal{E}(\rho_\theta)=\iE\left(\frac{1}{n}I + \sum_{i=1}^{n-1} \theta_i v_i\right)
=\frac{1}{n}I + \sum_{i=1}^{n-1}\lambda_{\pi_i} \theta_i v_i,
\end{equation}
where $\pi_i=k$, if $v_i \in \iA_k$.

\begin{ex}\label{ex2}
The simplest decomposition of $M_4$ to complementary subalgebras is 
$$
\iA_i=span\{ v_0, v_i \}, \quad i\in \{1,2,\dots, 15\}.
$$
Here we have subalgebras isomorphic to $\bbbc^2$, and from  (\ref{ch_gen}) the effect of the channel will be
$\mathcal{E}(\rho_\theta)=\frac{1}{4}I + \sum_{i=1}^{15}\lambda_{i} \theta_i v_i$.
\end{ex}

\begin{ex}\label{ex4}
Another decomposition of $M_4$ to complementary subalgebras: 
$$
\iA_1=span\{I,\sigma_0 \otimes \sigma_1,\sigma_0 \otimes \sigma_2,\sigma_0 \otimes \sigma_3\},
\iA_2=span\{I,\sigma_1 \otimes \sigma_0,\sigma_2 \otimes \sigma_0,\sigma_3 \otimes \sigma_0\}
$$
$$
\iA_3=span\{I,\sigma_1 \otimes \sigma_1,\sigma_2 \otimes \sigma_2,\sigma_3 \otimes \sigma_3\},
\iA_4=span\{I,\sigma_1 \otimes \sigma_2,\sigma_2 \otimes \sigma_3,\sigma_3 \otimes \sigma_1\}
$$
$$
\iA_5=span\{I,\sigma_1 \otimes \sigma_3,\sigma_3 \otimes \sigma_2,\sigma_2 \otimes \sigma_1\}.
$$
Let us remark that $\iA_1=\bbbc I \otimes M_2$ and $\iA_2= M_2 \otimes \bbbc I$ are isomorphic to $M_2$, while $\iA_3$, $\iA_4$ and $\iA_5$ are isomorphic to $\bbbc^4$.
The formula (\ref{ch_gen}) means heuristically, that we make contraction with constant $\lambda_j$ ($j=1,2,\dots 5$) in the $\iA_j$ direction, e.g., we multiply with $\lambda_1$ if $v_i=\sigma_0 \otimes \sigma_1$, or $v_i=\sigma_0 \otimes \sigma_2$, or $v_i=\sigma_0 \otimes \sigma_3$.
\end{ex}



\subsection{Parameter estimation}

If we want to estimate the channel parameters, first we send an input state  ($\rho_{\theta}$) through the channel, then perform a measurement $(M,I-M)$ on the channel output ($\iE(\rho_\theta)$), that results in an outcome related to $M$ with probability:
\begin{equation} \label{E:prob_ch}
p=\Tr(\mathcal{E}(\rho_\theta) M)=\frac{m_0}{\sqrt{n}}+\sum_{i=1}^{n-1} \lambda_{\pi_i} m_i  \theta_i 
\end{equation}
This shows that we can choose $\rho_{\theta}$ and $M$ as \textit{elements of the measurement configuration} 
to achieve an optimal estimate.

Let us suppose that we have $N$ identical copies of the input state. Having performed $N$ measurement on the output, we compute a relative frequency ($\nu$) of measuring $M$, which will be close to $p$  in (\ref{E:prob_ch}), and we can get an estimator $\hat\lambda_i$ 
that satisfies
\begin{equation} \label{E:est_ch}
\nu=\frac{m_0}{\sqrt{n}}+\sum_{i=1}^{n-1} \hat\lambda_{\pi_i} m_i  \theta_i 
\end{equation}
With $N_s$ unknown channel parameters we need $N_s$ input state -- measurement pairs, and so we will have $N_s$ equations of type (\ref{E:est_ch}), which can be solved uniquely.

The efficiency of estimation is usually characterized by the covariance matrix
\begin{equation}
\mathrm{Var}(\hat\lambda)_{i,j}= E(\hat\lambda_i \hat\lambda_j)- E(\hat\lambda_i)E( \hat\lambda_j).
\end{equation}
Another commonly used quantity for this is the Fisher information \cite{Petz-book}:
\begin{equation} \label{E:fisherdef}
F(\lambda)_{i,j}=\sum_\alpha \frac{1}{p_{\alpha}}
\frac{\partial p_{\alpha}}{\partial \lambda_i}
\frac{\partial p_{\alpha}}{\partial \lambda_j}
\end{equation}
Using Cramer-Rao inequality
\begin{equation}
\mathrm{Var}(\hat\lambda)\geq F(\lambda)^{-1},
\end{equation}
we can see that either maximizing Fisher information, or minimizing the variance can lead us to optimal parameter estimation.


\section{The qubit case with known channel directions }\label{sec:2dim_fix}

In this section we assume that the channel directions are known, and they are the Pauli directions, i.e. we have Eq. (\ref{eq:2x2out}) with $v_i=\sigma_i ~(i=1,2,3)$. The aim is to find an optimal measurement configuration, that consists of input state(s) and POVM that maximize the Fisher information of the estimated Pauli channel parameters. 

Let us suppose that,  $1 \ge |\lambda_1| \ge |\lambda_2| \ge |\lambda_3|$ and that the POVM related to $M$ form a von Neumann measurement, i.e. we have
\begin{equation}
M=1/2 \left(I+ \sum_{i=1}^3 m_i \sigma_i \right),
\end{equation}
with condition $m_1^2+m_2^2+m_3^2=1$.

We can calculate now the Fisher information matrix from (\ref{E:fisherdef}):
\begin{equation}\label{fishertr}
F(\lambda)_{i,j}=\frac{m_i \theta_i  m_j \theta_j}{1-\left(\sum_{i=1}^3 \lambda_i m_i \theta_i\right)^2}.
\end{equation}

We have three different parameters to estimate, so to have a general setting, let us suppose that we have three different measurements represented by the Bloch vectors  ($m^{(1)},m^{(2)},m^{(3)}$), with 3 different input states  ($\theta^{(1)},\theta^{(2)},\theta^{(3)}$). If we calculate the Fisher information matrix for each measurement ($F^{(j)}$), then we obtain the total information gained, $F^{\Sigma}=F^{(1)}+F^{(2)}+F^{(3)}$, by summing the three terms. 
We maximize the determinant of $F^{\Sigma}$, utilizing a straightforward extension of the Cramer-Rao inequality
\begin{equation}
\det(\mathrm{Var}(\hat\lambda))\ge  \det(F^{\Sigma}(\lambda)^{-1})=\det(F^{\Sigma}(\lambda))^{-1},
\end{equation}
The above form means that the volume of the covariance matrix will be minimal, which implies that we obtain a good estimate of all $\lambda_i$ parameters.

Using the notations $c_i^{(j)}=m_i^{(j)} \theta_i^{(j)}$ and $C_{i,j}=c_i^{(j)}$, the 
following form of the determinant is obtained 
\begin{equation} 
\det(F^{\Sigma}(\lambda))=\frac{\det(C)^2}{\prod\limits_{j=1}^3 \left[1-\left(\lambda_1 c_1^{(j)}+\lambda_2 c_2^{(j)}+\lambda_3 c_3^{(j)}\right)^2\right]}
\end{equation}
We should maximize this determinant as a function of $c^{(j)}$ with the proper conditions: $|c_1^{(j)}|+|c_2^{(j)}|+|c_3^{(j)}|\le 1$. 

\begin{theo}\label{th:2fix_2}
If we maximize the determinant of the Fisher information matrix with the above general measurement settings, then the optimal measurement configuration is 
$$\theta^{(1)}=(\pm1,0,0), ~m^{(1)}=(\pm 1,0,0),$$
$$\theta^{(2)}=(0,\pm1,0), ~m^{(2)}=(0,\pm 1,0),$$
$$\theta^{(3)}=(0,0,\pm1), ~m^{(3)}=(0,0,\pm 1).$$
\end{theo}
The proof is based on a modified gradient method (Mathematica \cite{math8} built-in optimizer). We choose the initial points as the elements on a grid of the state space. The result is robust, we get the same optimum starting from every input state - measurement pair.

Note that the same result was obtained in \cite{Ballo-2011} 
using convex maximization arguments. This approach does not seem to be well generalizable for the multidimensional case,
since even in two dimensions we have analytical problems. 

If we calculate the trace of Fisher-information, then Cramer-Rao inequality implies:
$$
\Tr(\mathrm{Var}(\hat\lambda))\ge  \Tr(F^{\Sigma}(\lambda)^{-1}) \ne \Tr(F^{\Sigma}(\lambda))^{-1}.
$$
The latter equality does not hold, so we should not maximize $\Tr(F^{\Sigma})$, instead let us check which 
input state -- measurement pair maximizes the trace of Fisher-information matrix (\ref{fishertr}). 

\begin{theo}\label{th:2fix}
If the trace of the Fisher information matrix is maximized, then the optimal 
input state is $\theta=(\pm1,0,0)$ and the optimal measurement is $m=(\pm 1,0,0)$, with $1 \ge |\lambda_1| \ge |\lambda_2| \ge |\lambda_3|$.
\end{theo}

\noindent
\textit{Proof:}
We need to maximize
\begin{equation}
\Tr F(\lambda) =\frac{\sum_{i=1}^3 m_i^2 \theta_i^2}{1-\left(\sum_{i=1}^3 \lambda_i m_i \theta_i\right)^2}.
\end{equation}
From Cauchy-Schwarz inequality we have
$$
|\sum_{i=1}^3 \lambda_i \theta_i \cdot m_i|^2 \le \left(\sum_{i=1}^3 \lambda_i^2 \theta_i^2\right) \left(\sum_{i=1}^3 m_i^2\right) \le \lambda_1^2  \left(\sum_{i=1}^3 \theta_i^2\right) \left(\sum_{i=1}^3 m_i^2\right)  \le \lambda_1^2
$$
and similarly 
$$
\sum_{i=1}^3 m_i^2 \theta_i^2\le \sqrt{\sum_{i=1}^3 \theta_i^4} \sqrt{\sum_{i=1}^3 m_i^4}\le  \sqrt{\sum_{i=1}^3 \theta_i^2} \sqrt{\sum_{i=1}^3 m_i^2} \le 1
$$
The inequalities are equalities with the given input state - measurement pairs, which implies the optimality of them.
\qed

It is important to note that in this case one obtains information about $\lambda_1$ only,  therefore, additional measurement configurations are needed for estimating the remaining two parameters. Using a greedy algorithm we can obtain the same result as in Theorem \ref{th:2fix_2}. 

\begin{cor}\label{cor}
If we want to measure all channel parameters, then the optimal setting is if we measure first $\lambda_1$ in the $\sigma_1$ direction, then $\lambda_2$ in the $\sigma_2$ direction, and finally $\lambda_3$ in the $\sigma_3$ direction. 
\end{cor}

\noindent
\textit{Proof:} Theorem \ref{th:2fix} ensures that the first measurement configuration should be $\theta=(\pm1,0,0),~m=(\pm1,0,0)$. In order to obtain information about the other channel directions, we should measure in the quasi-orthogonal subspace to get the highest possible information independently of the first measurement. This implies that $m_1=\theta_1=0$. We should notice, that the proof of Theorem \ref{th:2fix} does not depend on the actual length of the sums, so we have the same optimization problem except that there are fewer terms starting from $i=2$. The proof goes in the same way for that, too, hence the second optimal measurement setting is  $\theta=(0,\pm1,0),m=(0,\pm1,0)$, etc.
\qed

\medskip
In the latest scenario let us maximize the Fisher information for $\lambda_j$.

\begin{theo}\label{th:2j}
The maximal value of $F_{j,j}$ is taken when the optimal
input state and the optimal measurement are both in the $\sigma_j$ direction.
\end{theo}

\textit{Proof:}
We have to maximize the following expression
\begin{equation}
F_{j,j} =\frac{c_j^2}{1-\left(\sum_{i=1}^3 \lambda_i c_i\right)^2}\rightarrow \textrm{max}.
\end{equation}
Let us suppose that $j=3$, then we can write
$$
|\lambda_1 c_1+\lambda_2 c_2+\lambda_3 c_3| \le |\lambda_1| |c_1|+|\lambda_2| |c_2|+|\lambda_3| |c_3| \le |c_1|+|c_2|+|\lambda_3| |c_3|=
$$
$$
=(|c_1|+|c_2|+|c_3|)-|c_3|+|\lambda_3| |c_3|\le 1- (1-|\lambda_3|) |c_3|.
$$
We get that
$$
F_{3,3} \le I(|c_3|)=\frac{|c_3|^2}{1-\left( 1- (1-|\lambda_3|) |c_3|\right)^2}
$$
Simple calculation gives that $\partial I(|c_3|)/\partial|c_3|\ge 0$, so $I(|c_3|)$ is maximal if $|c_3|=1$, and we have $F_{3,3} \le \frac{1}{1-\lambda_3^2}$, and the equation is true for the input state and measurement given in the statement.
\qed

If we use the optimal input state -- measurement pairs for each $j$, we get here the same result as in the previous cases, too.


\section{Multidimensional case}

In this section our aim is to obtain a result similar to Section \ref{sec:2dim_fix} in higher dimension. 
First we notice that all the three cases of maximizing measures of the Fisher information matrix (determinant, trace, 
independently) resulted in the same optimal configuration for qubit Pauli channels, therefore we will use here the latest and 
simplest method presented in Theorem \ref{th:2j} for obtaining the most accurate estimation of $\lambda_j$. Thus, 
in the multidimensional case, we want to find the optimal input state (\ref{E:state_par}) and measurement 
(\ref{E:povm_par}) combination for which $F_{j,j}$ is maximized.

\subsection{The maximal value of the Fisher information}

We can calculate the Fisher-information matrix from (\ref{E:prob_ch}) directly 
\begin{equation}
F_{j,j}=\frac{c_j^2}{\left(\frac{m_0}{\sqrt{n}}+\sum\limits_{i=1}^{N_s}\lambda_i c_i\right)\left(1-\frac{m_0}{\sqrt{n}}-\sum\limits_{i=1}^{N_s} \lambda_i c_i\right)},
\end{equation}
using $c_i:=\sum_{\{l: ~v_l \in \iA_i\}} \theta_l m_l$, where we summarize the elements corresponding to $\lambda_i$. 

It is intuitively clear, that if we want to estimate $\lambda_j$, then we have to measure in the direction 
of $\lambda_j$, so we will suppose  $m \in \iA_j$ and $\theta \in \iA_j$. 

Then $c_i=0$, if $i\ne j$, so we get 
\begin{equation}\label{F_jj}
I(m_0,d):=F_{j,j}=\frac{d^2}{\left(\frac{m_0}{\sqrt{n}}+\lambda_j d\right)\left(1-\frac{m_0}{\sqrt{n}}-\lambda_j d\right)},
\end{equation}
where $d$ is defined in (\ref{eq:d}), and now $d=\sum c_i=c_j$. 
Calculating $\partial I(m_0,d)/\partial d$ we get, that the Fisher information is decreasing for negative $d$ values, and increasing for positive $d$ values, so it takes its maximum at the minimal or at the maximal value of $d$.
 Using the definition of $d$ we can obtain a general upper and lower bound:
\begin{equation}\label{E:bound_d}
0 \le p=\frac{m_0}{\sqrt{n}}+d \le 1 \quad  \Longrightarrow \quad d \in\left[-\frac{m_0}{\sqrt{n}},1-\frac{m_0}{\sqrt{n}}\right].
\end{equation}
We can assume that $m_0 \le \sqrt{n}/2$, since if $m_0>\sqrt{n}/2$ then we can use $I-M$ instead of $M$. 
If we substitute the lower bound $d=-m_0/\sqrt{n}$ into (\ref{F_jj}), we get that it is an increasing function of $m_0$, so it is maximal if $m_0=\sqrt{n}/2$. Using the maximal value $d=(1-m_0)/\sqrt{n}$, we have the same value for $F_{j,j}$ when  $m_0=\sqrt{n}/2$. Therefore, the Fisher information is maximal if the value of $d$ is maximal.

\medskip
We can use the Cauchy-Schwarz inequality to construct another estimation 
\begin{equation}\label{E:bound_d2}
|d+m_0 \theta_0|=| \sum_{i=0}^{n^2-1} m_i \theta_i | \le \sqrt{\sum_{i=0}^{n^2-1} m_i^2} \sqrt{\sum_{i=0}^{n^2-1} \theta_i^2}. 
\end{equation}

Furthermore we have the relationship
\begin{equation}\label{E:theta}
\fl \quad \sum_{i=0}^{n^2-1} \theta_i^2=\sum_{i,j} \theta_i \theta_j \Tr(v_i \cdot v_j)=\Tr\left(\sum_i \theta_i v_i \cdot \sum_j \theta_j v_j\right)=\Tr \rho^2=\sum_{k=1}^{n} \mu_k^2,
\end{equation}
where $\mu_k$-s are the eigenvalues of $\rho$. 
In addition, we know that $\mu_k \ge 0$ and $\sum_{k=1}^{n} \mu_k =1$. 

Assume that the eigenvalues have multiplicity $K$, then it is easy to obtain that
\begin{equation}\label{ineq_mu}
\sum_{k=1}^{n} \mu_k^2\le \frac{1}{K}.
\end{equation}

We can make a similar upper bound for the $m_i$-s in (\ref{E:bound_d2}), thus we get the general upper bound
\begin{equation}\label{E:bound_d3}
|d+\frac{m_0}{\sqrt{n}}| \le \frac{m_0\sqrt{n}}{K}  \Longrightarrow \quad d \le -\frac{m_0}{\sqrt{n}}+\frac{m_0\sqrt{n}}{K}
\end{equation}
It is easy to check, that the Fisher information is increasing in $m_0$ if $d=\frac{m_0\sqrt{n}}{K}-\frac{m_0}{\sqrt{n}}$ (upper bound from (\ref{E:bound_d3})), while  the Fisher information is decreasing in $m_0$ if $d=1-\frac{m_0}{\sqrt{n}}$  (upper bound from (\ref{E:bound_d})). 
Therefore, the Fisher information is maximal if the two upper bounds are equal, 
hence $m_0=\frac{K}{\sqrt{n}}$ and $d=\frac{n-K}{n}$ and we can conclude the following statement:

\begin{theo}\label{Th:n}
If $m \in \iA_j$ and $\theta \in \iA_j$ and quantum states in $\iA_j$ have multiplicity $K$, then the maximal achievable
Fisher information is
\begin{equation}\label{Imax}
I_{max}=\frac{1}{(1-\lambda_j)(\frac{K}{n-K}+\lambda_j)}.
\end{equation}
\end{theo}

We can use the above result to obtain some general statements. 

\begin{cor}\label{Th:C2k}
If $\iA_j$ is isomorphic to $\bbbc^{k} ~(k \ge 2)$, then the maximal Fisher information is 
$$I_{max}^{(M_n~:~\bbbc^k)}=\frac{1}{(1-\lambda_j)(\frac{1}{k-1}+\lambda_j)}.$$
\end{cor}

\noindent
\textit{Proof:}
If $\iA_j$ is isomorphic to $\bbbc^k$ then the eigenvalues are all with multiplicity $K=n/k$, and Theorem \ref{Th:n} implies 
the result.
\qed

\begin{cor}\label{Th:M2k}
If $\iA_j$ is isomorphic to $M_{m} ~(m \ge 2)$, then the maximal Fisher information is 
$$I_{max}^{(M_n~:~M_m)}=\frac{1}{(1-\lambda_j)(\frac{1}{m-1}+\lambda_j)}.$$
\end{cor}

\noindent
\textit{Proof:}
If $\iA_j$ is isomorphic to $M_{m}$ then the eigenvalues are all with multiplicity $K=n/m$, and Theorem \ref{Th:n} implies 
the result.
\qed

\begin{ex}
Using the decomposition in Example \ref{ex2} we can get the maximal Fisher information: 
$$I_{max}^{(M_4~:~\bbbc^2)}=\frac{1}{(1-\lambda_j)(1+\lambda_j)}, \quad \forall j.$$
\end{ex}

\begin{ex}
Using the decomposition in Example \ref{ex4} we can get the maximal Fisher information: 
$$I_{max}^{(M_4~:~M_2)}=\frac{1}{(1-\lambda_j)(1+\lambda_j)}, \quad\textrm{for}\quad j\in \{1,2\}$$
and
$$I_{max}^{(M_4~:~\bbbc^4)}=\frac{1}{(1-\lambda_j)(\frac13+\lambda_j)}, \quad\textrm{for}\quad j\in \{3,4,5\}.$$
\end{ex}

We can see that the maximal Fisher information strongly depends on the algebraic structure of $\iA_j$-s even in this simple scenario, 
and for the same $\lambda_j$ values we can get more information if $\iA_j$ is isomorphic to $\bbbc^4$ than if $\iA_j$ is isomorphic 
to $M_2$ or $\bbbc^2$, since $I_{max}^{(M_4~:~\bbbc^4)}>I_{max}^{(M_4~:~M_2)}=I_{max}^{(M_4~:~\bbbc^2)}$.

\subsection{The optimal measurement configuration}

It is in general hard to give an optimal input state -- measurement pair, since it will depend on the structure of $\iA_j$, but we can 
characterize them. 

To obtain upper bound (\ref{E:bound_d3}), we used 2 inequalities, so if we want to find the optimal measurement settings,
both inequalities should be satisfied with an equality. In (\ref{E:bound_d2}) the Cauchy-Schwarz inequality is sharp
if $m$ and $\theta$ are multiples of each other. On the other hand,
the maximum is achieved in (\ref{ineq_mu}) when we have $\mu_k=\frac{1}{K}$ for $K$ different $k$ value, and 
$\mu_k=0$ in the other cases.
From this and from $m_0=\frac{K}{\sqrt{n}}=K\cdot \theta_0$ follows, that the optimal measurement ($m_{opt}$) is an arbitrary von Neumann measurement in $\iA_j$ with 
minimal rank ($K$), and the optimal input state is $\theta_{opt}=\frac{ m_{opt}}{K}$. 
 
Let us also note that the proof works either for positive and negative $\lambda_j$, however there is some asymmetry: 
the maximum value is $\lambda_j\le 1$, but from (\ref{eq:genPauli_inequ}) we get $\lambda_j\ge -\frac{1}{n-1}$.


\section{Unknown channel directions in the qubit case}

In the previous sections we obtained results for known, given channel directions. 
In this section we propose a method to estimate the direction of axes too ($\varphi_i$), beside the magnitude of contraction ($\lambda_i$). 
Obtaining the optimal measurement scenario is much harder here, therefore the efficiency of the proposed algorithm is only investigated using simulations.

\subsection{The estimation scheme}

In the qubit case, the quantum states can be described with Bloch-vectors (\ref{E:2x2}), hence they can be considered as elements of the $\R^3$ space.
Therefore, the effect of the Pauli channel can be described with a linear operation on this space in the following way:
\be
\ep(\rho_\theta)=\rho_{\theta^*}, \quad  \textrm{iff} \quad \theta^*=A \theta,
\label{efct}
\ee
where $A$ denotes the appropriate linear operator. 
\par
The channel has six real parameters: three angles ($\varphi_1, \ \varphi_2, \ \varphi_3 $) and three eigenvalues 
($\lambda_{1}, \ \lambda_{2},\ \lambda_{3}$); the angles determine the channel directions, 
the eigenvalues are the parameters of the contraction. We will use the following decomposition of $A$:
\be
\label{alap}
A( \lambda_1,\lambda_2, \lambda_3,\varphi_1, \varphi_2,  \varphi_3)=R_{z} R_{y} R_{x} \Lambda R_{x}^{-1} R_{y}^{-1} R_{z}^{-1}
\ee
where
\[
R_{z} (\varphi_1)=
\left(
\begin{array}{ccc}
\cos{\varphi_1} & -\sin{\varphi_1} & 0 \\
\sin{\varphi_1} & \cos{\varphi_1} & 0 \\
0 & 0 & 1 
\end{array}
\right), 
\
R_{y} (\varphi_2)=
\left(
\begin{array}{ccc}
\cos{\varphi_2} & 0 & -\sin{\varphi_2} \\
0 & 1 & 0 \\
\sin{\varphi_2} & 0& \cos{\varphi_2}
\end{array}
\right),
\]
\[
R_{x} (\varphi_3)=
\left(
\begin{array}{ccc}
1 & 0 & 0\\
0 & \cos{\varphi_3} & -\sin{\varphi_3} \\
0 & \sin{\varphi_3} & \cos{\varphi_3}
\end{array}
\right)
\quad \textrm{and} \quad 
\
\Lambda=
\left(
\begin{array}{ccc}
\lambda_{1} & 0 & 0\\
0 & \lambda_{2} & 0 \\
0 & 0 & \lambda_{3}
\end{array}
\right).
\]

We choose three input qubits, such that their Bloch-vectors
($\ul{\theta}^{(1)}, \ul{\theta}^{(2)}, \ul{\theta}^{(3)}$) are linearly independent. 
Using notation $\ul{\ul{\theta}}=[\ul{\theta}^{(1)}, \ul{\theta}^{(2)}, \ul{\theta}^{(3)}]\in M_3(\R)$, we have the 
following equation for output qubits:
\be\label{output}
\ul{\ul{\theta}}^*=A \cdot \ul{\ul{\theta}}
\ee
\par
Thereafter, we estimate the output qubits of the channel with three projective measurements: $(M^{(i)},I-M^{(i)}),\quad i=\{1,2,3\} $, where $M^{(i)}=\frac{1+ \ul m^{(i)}}{2}$. If we denote the probability of the outcome of measuring $M^{(i)}$ for the input state $\ul{\theta}^{(j)}$ by $\ul{\ul{P}}_{i,j}$, and using the notation $\ul{\ul{ M}}=[\ul m^{(1)},\ul m^{(2)},\ul m^{(3)}]$, we obtain 
\be\label{prob_dir}
\ul{\ul{P}}=\frac{1+ \ul{\ul{ M}}^T \cdot \ul{\ul{\theta}}^*}{2}.
\ee
Let $N_{i,j}$ denote the number of measurements
corresponding to  $\ul{m}^{(i)}$ and $\ul{\theta}^{(j)}$, and $N_{i,j}(+)$ the number of $M^{(i)}$ outcomes, then the relative frequency $\nu_{i,j}=\frac{N_{i,j}(+)}{N_{i,j}}$ will be an unbiased estimator for $\ul{\ul{P}}_{i,j}$.
Then using (\ref{prob_dir}) we obtain an estimation on $\ul{\ul{\theta}}^*$
\be\label{kalap}
\hat{\ul{\ul{\theta}}^*}=\left(\ul{\ul{M}}^T\right)^{-1} (2\nu-1),
\ee
where $\nu$ is the matrix with elements $\nu_{i,j}$, and since it is linear in $\nu$ we will get an unbiased estimation for $\ul{\ul{\theta}}^*$, too.

From (\ref{output}) we can estimate the channel effect in the form 
\be
\hat A=\ul{\ul{\hat\theta}}^*\cdot \ul{\ul{\theta}}^{-1}
\ee
The estimated channel effect determines the parameter estimations (\ref{alap}),
so we can construct equations for determining the channel parameters in the form 
\be\label{equ}
A(\hat \lambda_1, \hat \lambda_2, \hat \lambda_3, \hat \varphi_1, \hat \varphi_2, \hat \varphi_3)=\frac{\hat A+\hat A^T}{2}.
\ee
We use symmetrization of $\hat A$ since we know that A is symmetric. 
Therefore, it is enough to use 
only the upper triangular part of $A$ for parameter estimation, 
because there are only 6 parameters. 

By solving the matrix equation $(\ref{equ})$ we get the estimation on the channel parameters.

\subsection{The efficiency of the estimation}

The parameters of the Pauli channel are fixed and we can consider the input qubits and measurements on the output qubits as optimization variables. 
Our aim is to find the optimal measurement setup by solving the minimization problem:
\be\label{minim}
\mathcal{V}=(1-c)\left(\sum_{i=1}^3\Var(\hat\lambda_i)\right)+c \left(\sum_{i=1}^3 \Var(\hat\varphi_i)\right) \rightarrow \textrm{ min}.
\ee
A constant $c\in[0,1]$ is used here to express our preference with respect to the 
two sets of parameters. With $c=1$ we can minimize the estimation error of the axis 
 direction, and with $c=0$  
we can minimize the error of the contraction magnitudes, while by using $0<c<1$ we can minimize them simultaneously.

The problem is that the equations in (\ref{equ}) are non-linear in angle parameters, so the estimators will not be unbiased even if we know that $\hat A$ is unbiased. 
Furthermore, the variances in (\ref{minim}) can not be expressed analytically.

Therefore, one may investigate the solution of the above optimization problem by 
 using simulations. In this case, we generate $K$ realizations of the random 
 variable $\nu$ for a given $\ul{\ul{ M}}$ and $\ul{\ul{\theta}}$, using the built-in random function of Mathematica \cite{math8}. From each realization of $\nu$ we get an estimate for $\hat A$, and by solving equation (\ref{equ}) we can estimate the channel parameters, and from that we can get the empirical mean square error (MSE): 
\be\label{MSE}
\widehat{MSE(\hat{\vartheta})}= \frac{1}{K}\sum_{k=1}^{K} \left(\hat{\vartheta}_k-\vartheta \right)^2,
\ee
where $\vartheta$ is one of the estimated parameters ($\vartheta=\lambda_i~ \mathrm{ or } ~\varphi_i$) and $\hat \vartheta_k$ is the solution that is obtained from (\ref{equ}) for the $k$-th realization.
Note that in \eref{MSE} the mean square error is used instead of the variance. 
We know that
\be
MSE(\hat{\vartheta})=\Var(\hat{\vartheta})+(\bE(\hat{\vartheta})-\vartheta)^2
\ee
so aside from showing that the variance vanishes, the MSE tending to zero indicates that the estimation method is asymptotically unbiased.

By substituting the quantities \eref{MSE} into (\ref{minim}) instead of the variances, we obtain an estimate $\hat\mathcal{V}$, which is a good approximation of $\mathcal{V}$ if $K$ is large enough $\left(\widehat{MSE(\hat{\vartheta})}\rightarrow MSE(\hat{\vartheta})\right)$ and the bias of $\hat \vartheta_k$ is small (which turns out to be true for large $N_{i,j}$-s).

\subsection{Simulation results}

For the simulation investigations the following constant system parameters were used: 
$\lambda_1=3/4$, $\lambda_2=1/2$, $\lambda_3=1/4$, $\varphi_i=0$, for $i \in \{1,2,3\}$. 

In each direction the same number of measurements $N_{i,j}=N$ were used, and we repeated the whole measurement process many times ($K=1000$) to ensure, that the values of $\hat\mathcal{V}$ are close to the real mean square error. 


\begin{figure}[!ht]
\begin{center}$
  \begin{array}{cc}
   \includegraphics[width=7.5cm]{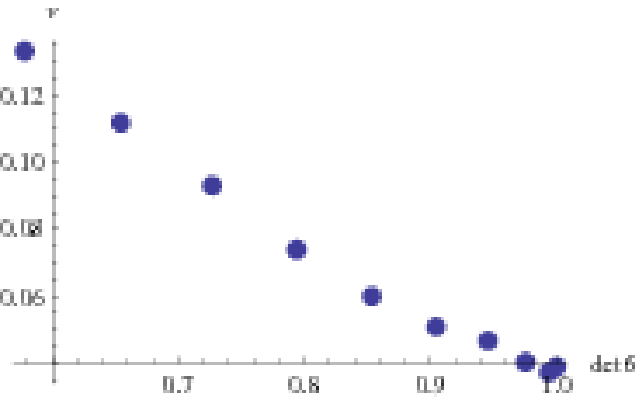}&
    \label{fig:orthogonal}
  \includegraphics[width=7.5cm]{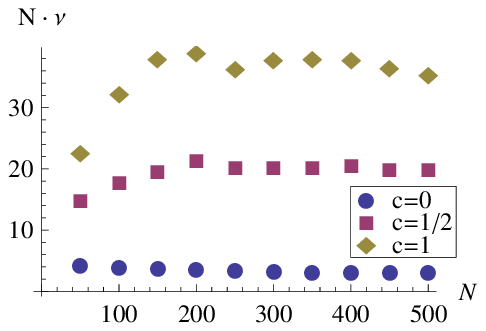}\label{fig:c_value}
     \\(a)&(b)
    \end{array}$
\caption{ (a) $\hat\mathcal{V}$ as a function of $\det\left (\ul{\ul{\theta}}\right)$, (b) $N \cdot \hat\mathcal{V}$ for difference $c$ values as a function of number of measurements. \label{fig1}}
\end{center}
\end{figure}

First we remark that $\ul{\ul{ M}}$ and $\ul{\ul{\theta}}$ are invertible, and we used their inverses in our calculations. Moreover, the results with known channel directions reported in the earlier sections suggest that they are probably orthogonal in the optimal case. Therefore,  $\hat\mathcal{V}$ was investigated as a function of the input state triplets 
with a randomly chosen but fixed $M$ measurement matrix. For this purpose, some pure states were used as input states (the angle between them was gradually increased), and the value $\det\left (\ul{\ul{\theta}}\right)$ was used as a measure of orthogonality. 
The result of this investigation can be seen in Figure \ref{fig1}(a), where it is visible that smaller variance was obtained for the orthogonal case. 
The same result was observed when a fixed input state triplet was chosen and the 
measurement directions were changed: the orthogonal measurement direction case produces the smallest variance. 
As a conclusion, \textit{orthogonal matrices for $\ul{\ul{ M}}$ and $\ul{\ul{\theta}}$ were found to be optimal}.

Thereafter we investigated the variance for three different $c$ values: $c=0,~ c=0.5$ and $c=1$. 
For this purpose, a fixed random orthogonal measurement direction set, and input state triplet $\ul{\ul{\theta}}=I$ were used. The result is shown in Figure \ref{fig1}(b). It is seen,  that $N \cdot \hat\mathcal{V}$ is nearly constant for all $c$ values (except for low $N$ values), so the MSE converges to 0 in $1/N$ order. This means that we can obtain accurate state estimation for all cases if $N$ is large enough. The difference is in the coefficients: if $c=0$ then the variance is smaller, so we can estimate $\lambda$ more accurately than the angles from the same number of measurements. This is not surprising, since then the estimators are linear. 

A further useful observation is, that for every $0<c<1$ the variance can be expressed as a superposition of the $c=0$ and $c=1$ cases, so we can separate the problem into two parts: to measuring the directions of the channel and to measuring the magnitude of contractions. 
The latter has been discussed in Section \ref{sec:2dim_fix} and an analytical solution was found, so it is enough to examine the $c=1$ case by using simulation.

Section \ref{sec:2dim_fix} would suggest that the optimal input states and measurements should be in the channel directions. However, this is not true in the $c=1$ case: by rotating $M$ we could achieve a better value for $\hat\mathcal{V}$ than with $\ul{\ul{ M}}=\ul{\ul{\theta}}=I$. Therefore, further investigations are needed to find the optimal input states and measurements in the $c=1$ case, i.e.\ when we want to estimate the channel directions.


\section{Discussion and Conclusions}

In this article we examined different scenarios to find optimal parameter estimation schemes for Pauli channels, that consist of the optimal input state and the optimal two-element POVM.

Assuming known channel directions, both the qubit and the $n$-dimensional cases were discussed in detail.
In the two-dimensional case we have shown that the optimal measurement-input state pairs should be directed towards the channel directions. 
We have achieved the same optimal result in three different ways: first we gave a jointly optimal estimation for the parameters, 
then we maximized the total achievable information in each step, and finally we constructed the best estimation for $\lambda_j$-s independently.

In the $n$-dimensional case the independent $\lambda_j$ estimation method was used because of its simplicity. 
We have obtained similar results that in the qubit case: in the optimal estimation scheme of $\lambda_j$ the 
measurement and input states are in the corresponding subalgebra $\iA_j$, and they are multiples of each other.
 Furthermore, the measurement should be a von Neumann measurement with minimal rank in $\iA_j$. This implies that 
the optimal Fisher information depends on the algebraic structure of $\iA_j$.

Finally, we investigated the case when we have to estimate the channel directions, too. The quantity (\ref{minim}) was used to measure the quality of estimation, and the estimates were obtained by its minimization. A tuning parameter was introduced to weight the variance of  channel direction estimations to that of contractions. It was found that to determine analytically the optimal estimation scheme of channel directions is hard even in the qubit case. 
Our simulation investigations show that the mean square error of the proposed method converges to zero with the order of $1/N$, where $N$ is the number of used measurements, so we can achieve the same estimation order as in $\lambda_j$ case in spite of non-linearity. Our simulations also indicate that we do not necessarily achieve the smallest variance if we measure in the channel direction.

\ack This research was supported in part by the Hungarian Research Fund through grant $K83440$. 

\section*{References}

\bibliographystyle{plain}
\bibliography{pauli-quantum}

\begin{thebibliography}{10}

\bibitem{Ballo-2011}
G.~Ball\'o and K.~M. Hangos.
\newblock Experiment design and parameter estimation of {P}auli channels using
  convex optimization.
\newblock {\em Preprint: arxiv.org/1107.0890}, 2011.

\bibitem{Bendersky-2008}
A.~Bendersky, F.~Patawski, and J.P. Paz.
\newblock Selective and efficient estimation of parameters for quantum process
  tomography.
\newblock {\em Phys. Rev. Lett.}, 100:190403, 2008.

\bibitem{Branderhorst2009}
M.P.A. Branderhorst, J.~Nunn, I.A. Walmsley, and R.L. Kosut.
\newblock Simplified quantum process tomography.
\newblock {\em New Journal of Physics}, 11:115010, 2009.

\bibitem{Bschorr-2001}
T.~C. Bschorr, D.~G. Fischer, and M.~Freyberger.
\newblock Channel estimation with noisy entanglement.
\newblock {\em Physics Letters A}, 292:15--22, 2001.

\bibitem{Chiuri-2011}
A.~Chiuri, V.~Rosati, G.~Vallone, S.~Padua, H.~Imai, S.~Giacomini,
  C.~Macchiavello, and P.~Mataloni.
\newblock Experimental realization of optimal noise estimation for a general
  {P}auli channel.
\newblock {\em Preprint: arxiv.org/1107.0888}, 2011.

\bibitem{Dariano-2003}
G.~M. D'Ariano, M.~G.~A. Paris, and M.~F. Sacchi.
\newblock Quantum tomography.
\newblock {\em Advances in Imaging and Electron Physics}, 128:205, 2003.

\bibitem{Dariano-2005}
G.~M. D'Ariano, M.~F. Sacchi, and J.~Kahn.
\newblock Minimax discrimination of two {P}auli channels.
\newblock {\em Phys. Rev. A}, 72:052302, 2005.

\bibitem{Fujiwara2003}
A.~Fujiwara and H.~Imai.
\newblock Quantum parameter estimation of a generalized {P}auli channel.
\newblock {\em Jornal of Physics A: Mathematical and General}, 36:8093--8103,
  2003.

\bibitem{Ji-2008}
Z.~Ji, G.~Wang, R.~Duan, Y.~Feng, and M.~Ying.
\newblock Parameter estimation of quantum channels.
\newblock {\em IEEE Trans. Inf. Theory}, 54:5172--5185, 2008.

\bibitem{Mohseni-2008}
M.~Mohseni, A.~T. Rezakhani, and D.~A. Lidar.
\newblock Quantum process tomography: Resource analysis of different
  strategies.
\newblock {\em Physical Review A}, 77:032322, 2008.

\bibitem{Nathanson-2007}
M.~Nathanson and M.~B. Ruskai.
\newblock Pauli diagonal channels constant on axes.
\newblock {\em Journal of Physics A Mathematical General}, 40:8171--8204, 2007.

\bibitem{Nielsen-book}
M.~A. Nielsen and I.~L. Chuang.
\newblock {\em Quantum Computation and Quantum Information}.
\newblock {Cambridge University Press}, 2000.

\bibitem{Nunn2010}
J.~Nunn, B.~J. Smith, G.~Puentes, I.~A. Walmsley, and J.~S. Lundeen.
\newblock Optimal experiment design for quantum state tomography: Fair,
  precise, and minimal tomography.
\newblock {\em Physical Review A}, 81:042109, 2010.

\bibitem{Petz-book}
D.~Petz.
\newblock {\em Quantum Information Theory and Quantum Statistics}.
\newblock Theoretical and Mathematical Physics. Springer-Verlag, 2008.

\bibitem{Petz-2008}
D.~Petz and H.~Ohno.
\newblock Generalizations of {P}auli channels.
\newblock {\em Acta Math. Hungar}, 124:165--177, 2009.

\bibitem{Sarovar2006}
M.~Sarovar and G.J. Milburn.
\newblock Optimal estimation of one-parameter quantum channels.
\newblock {\em Journal of Physics A: Mathematical and General}, 39:8487, 2006.

\bibitem{Sasaki-2002}
M.~Sasaki, M.~Ban, and S.~M. Barnett.
\newblock Optimal parameter estimation of a depolarizing channel.
\newblock {\em Phys. Rev. A}, 66(2):022308, 2002.

\bibitem{math8}
Inc. Wolfram~Research.
\newblock {\em Mathematica Edition: Version 8.0}.
\newblock Wolfram Research, Inc., Champaign, Illinois, 2010.

\bibitem{Young-2009}
K.~C. Young, M.~Sarovar, R.~Kosut, and K.~B. Whaley.
\newblock Optimal quantum multiparameter estimation and application to dipole-
  and exchange-coupled qubits.
\newblock {\em Phys. Rev. A}, 79(6):062301, 2009.

\bibitem{Ziman-2005}
M.~Ziman, M.~Plesch, V.~Buzek, and P.~Stelmachovic.
\newblock Process reconstruction: From unphysical to physical maps via maximum
  likelihood.
\newblock {\em Phys. Rev. A}, 72:022106, 2005.

\end{thebibliography}

\end{document}